\newcounter{myctr}
\def\myitem{\refstepcounter{myctr}\bibfont\noindent\ifnum\themyctr>9\else\phantom{0}\fi\hangindent17pt\themyctr.\enskip}

\documentclass{ws-ijqi}

\begin{document}

\markboth{Authors' Names}
{Instructions for Typing Manuscripts (Paper's Title)}

%%%%%%%%%%%%%%%%%%%%% Publisher's Area please ignore %%%%%%%%%%%%%%
\catchline{}{}{}{}{}
%%%%%%%%%%%%%%%%%%%%%%%%%%%%%%%%%%%%%%%%%%%%%%%%%%%%%%%%%%%%%%%%%%%

\title{ENTANGLEMENT AND OTHER QUANTUM CORRELATIONS\\
OF A SINGLE QUDIT STATE\\ AS A RESOURCE FOR QUANTUM TECHNOLOGIES }

\author{MARGARITA A. MAN'KO}

\address{P.N.~Lebedev Physical Institute, Russian Academy of Sciences\\
Leninskii Prospect 53, Moscow 119991, Russia\\
mmanko@sci.lebedev.ru}

\author{VLADIMIR I. MAN'KO}

\address{P.N.~Lebedev Physical Institute, Russian Academy of Sciences\\
Leninskii Prospect 53, Moscow 119991, Russia\\
manko@sci.lebedev.ru}

\address{Moscow Institute of Physics and Technology (State University)\\
Institutski\'{\i} per. 9, Dolgoprudny\'{\i}, Moscow Region 141700,
Russia}

\maketitle

\begin{history}
\received{Day Month 2014}
\revised{Day Month 2014}
%\accepted{Day Month Year}
%\comby{(xxxxxxxxxx)}
\end{history}

\begin{abstract}
The approach to extend the notion of entanglement for characterizing
the properties of quantum correlations in the state of a single
qudit is presented. New information and entropic inequalities, such
as the subadditivity condition, strong subadditivity condition, and
monotonicity of relative entropy for a single qudit corresponding to
an arbitrary spin state with spin $j$ are discussed. The idea to
employ quantum correlations in the single-qudit state, such as the
entanglement, for developing a new quantum technique in quantum
computing and quantum communication is proposed. Examples of qutrit
and qudit with $j=3/2$ are considered.
\end{abstract}


\section{Introduction}

The entanglement phenomenon~\cite{Schroed35} is associated with
quantum correlations in composite systems containing subsystems.
Entangled states of the composite systems, e.g., of several qubits
are considered as a resource for their employment in quantum
technologies like quantum computing, quantum teleportation, and
quantum coding~\cite{Chuang}.

Quantum correlations of composite systems are also characterized by
the discord~\cite{discord1,discord2,YurkevichJRLR}. The notion of
discord is associated with information properties of bipartite
quantum system, e.g., two qubits. The bipartite system states are
determined by the density operator matrix $\rho(1,2)$, which acts in
the Hilbert space $H$ of the system states, and this Hilbert space
has the structure of the tensor product $H=H_1\otimes H_2$ of the
Hilbert spaces of the first and second subsystem states.

Bipartite and multipartite quantum system states have entropic and
information characteristics. The von Neumann entropy of the
bipartite system state satisfies specific inequalities, which are
numerical inequalities for matrix elements of the density matrices
of the system and its subsystems.

The aim of our work is to show that the inequalities for the density
matrices analogous to the known
inequalities~\cite{Lieb,Petz,Carlen1,Carlen2,20pr,20pra,20prd,20pre,Lieb2014,Zhang}
for the composite systems can be obtained also for non-composite
systems, e.g., for a single qudit. Also we show that there exist the
entanglement properties of the density matrix of the single qudit
state, and the notion of mutual information can be extended and
applied to the single qudit state. This means that the resource of
quantum correlations, which are considered from the viewpoint of the
possibility to be applied in quantum technologies, is available not
only in composite systems as correlations of their subsystems but in
non-composite systems as well.

In this paper, we present some new inequalities following the
approach suggested in
\cite{PS-MA-VI-1,PS-MA-VI-2,OlgaVovaVIminkovski,OlgaVova,Markovich,OlgaVova1}.
It is worth mentioning that our approach is coherent with the
approach to quantum correlations in the singe qudit state discussed
in
\cite{Shumovsky}. Also some aspects of our approach are close to
the consideration of Laplace matrices in
\cite{Braunstein,Mancini,Wu}.

This paper is organized as follows.

In section~2, we obtain new entropic inequalities on an example of
the qutrit state. We present the inequalities for tomograms of qudit
states in section~3 and discuss the monotonicity of the
single-qudit-state density matrices in section~4. We find the Bell
inequality for the $j=3/2$ single qudit in section~5 and give our
conclusions and prospectives in section~6.

\section{Information and entropic inequalities}

Before formulating the information and entropic inequalities, first
we study linear maps of matrices. Following~\cite{SudarJRLR}, we
consider rectangular matrices as vectors. Let matrix $a$ have matrix
elements $a_{jk}$, where $j=1,2,\ldots,n$ and $k=1,2,\ldots,m$. We
construct the linear map $\hat Va=\vec a$, where the $N$-vector
$\vec a$ has $nm$ components, i.e., $\vec
a=(a_{11},a_{12},\ldots,a_{1m},a_{21},a_{22},\ldots,a_{2m},\ldots,a_{n1},a_{n2},\ldots,a_{nm})$.

The inverse map provides a matrix from the $N$-vector, i.e., $a=\hat
V^{-1}\vec a$.

Using the vectors $\vec a$ instead of the matrices $a$, one can
describe the linear maps of the matrices by means of linear
transforms of the corresponding vectors $\vec a$. This means that
the map of matrix $a\to b$ denoted as
\begin{equation}\label{1.1}
b=\hat La
\end{equation}
has the matrix form
\begin{equation}\label{1.2}
b_J=\sum_{K=1}^NL_{JK}a_K.
\end{equation}
We consider the case of $m=n$ and introduce some special transforms
of the matrix $a$ as follows:
\begin{equation}\label{1.3}
b=P_{n-1}aP_{n-1}+P_{n}aP_{n}=\hat La,
\end{equation}
where the $n$$\times$$n$-matrix $P_n$ acts on any $n$-vector $\vec
c=(c_1,c_2,\ldots,c_n)$ as a projector of the form $P_n\vec
c=(0,0,\ldots,c_n)$. The matrix $P_{n-1}=1_n-P_n$, where $1_n$ is
the identity $n$$\times$$n$-matrix, and the matrix $L_{JK}$ in this
case has some nonzero diagonal matrix elements equal to unity.
Non-diagonal matrix elements are equal to zero. The zero diagonal
matrix elements read
\begin{equation}\label{1.4}
L_{JK}=\delta_{JK}f(J),
\end{equation}
where the function $f(J)$ is equal to zero for the following values
of the arguments
\begin{equation}\label{1.5}
f(n)=f(2n)=\cdots=f((n-1)n)=f(n^2-n+1)=f(n^2-n+2)=\cdots=f(n^2-1)=0.
\end{equation}
All the other values of the function $f(J)=1$.

We denote the matrix $L_{JK}$~(\ref{1.4}) with the properties
described by equation~(\ref{1.5}) as a matrix $M_2$. For example, if
$n=3$, the 9$\times$9-matrix $M_2$ has the block form
\begin{equation}\label{1.6}
M_2=\left(
          \begin{array}{ccc}
            {\cal P}_2 & 0 & 0 \\
            0 & {\cal P}_2 & 0\\
            0 & 0 & {\cal P}_1
          \end{array}
        \right);  \qquad
        {\cal P}_2 =\left(
          \begin{array}{ccc}
            1 & 0 & 0\\
            0 & 1 & 0 \\
            0 & 0 & 0
          \end{array}
        \right);\qquad
        {\cal P}_1 =\left(
          \begin{array}{ccc}
            0 & 0 & 0\\
            0 & 0 & 0 \\
            0 & 0 & 1
          \end{array}
        \right).
\end{equation}

If the matrix $a$ is the density matrix, i.e., $a=a^\dagger$,
Tr$\,a=1$, and $a\geq 0$, the map given by equation~(\ref{1.3}) is
the positive map. In terms of the matrix $M_2$, this map corresponds
to the transform of the vector $\vec a$, i.e., $\hat V^{-1}(M_2\vec
a)$ is the matrix with the properties of a qutrit density matrix.
The $N$$\times$$N$-matrix $L_{JK}$~(\ref{1.4}) can be written in the
form
\begin{equation}\label{1.7}
 L_{JK}=\big(P_{n-1}\otimes P_{n-1}+P_{n}\otimes P_{n}\big) _{JK}.
\end{equation}
This formula is the partial case to obtain the matrix $L_{JK}$,
which corresponds to the linear transform of the matrix $a$ given in
the form
\begin{equation}\label{1.8}
a\to b=\hat La=\sum_sP_saP_s^\dagger,
\end{equation}
where $P_s$ are arbitrary matrices. In this case, the matrix
$L_{JK}$ reads
\begin{equation}\label{1.9}
 L_{JK}=\Big(\sum_sP_s\otimes P_s^*\Big) _{JK}.
\end{equation}
The index $s$ in (\ref{1.8}) and (\ref{1.9}) can take values in an
arbitrary domain. In the case where the matrices $P_s$ are
projectors, formula~(\ref{1.9}) provides the diagonal matrix
$L_{JK}$.

For matrices $a$, which are the density matrices, we denote the
matrix $L_{JK}$ as $M_2$. The transform~(\ref{1.9}) with projectors
$P_s$ yields the positive map (completely positive map) of the
matrix $a$. The positive transform of the density matrix can be
interpreted as a decoherence transform, which corresponds to
equating some non-diagonal matrix elements of the density matrix to
zero.

One can construct another map of the matrix $a$. This map is given
by the transform of the vector $\vec a\to M_1\vec a$, where the
matrix $M_1$ reads
\begin{equation}\label{1.10}
M_1=\left(
          \begin{array}{ccc}
            \Pi_1 & 0 & 0 \\
            0 &  \Pi_2 & 0\\
            0 & 0 & \Pi_1
          \end{array}
        \right);  \qquad
        \Pi_1 =\left(
          \begin{array}{ccc}
            1 & 0 & 0\\
            0 & 0 & 0 \\
            0 & 0 & 1
          \end{array}
        \right);\qquad
        \Pi_2 =\left(
          \begin{array}{ccc}
            0 & 0 & 0\\
            0 & 1 & 0 \\
            0 & 0 & 0
          \end{array}
        \right).
\end{equation}

One can construct another matrix $\widetilde M_2$ of the form
\begin{equation}\label{1.11}
\widetilde M_2=M_2+S_1-P,
\end{equation}
where the only nonzero matrix element in $N$$\times$$N$-matrices
$S_1$ and $P$ is
\begin{equation}\label{1.12}
(S_1)_{1N}=P_{NN}=1.
\end{equation}
Then for the matrix $a$ one has the map
\begin{equation}\label{1.13}
a\to\tilde a_2=\hat V^{-1}\big(\widetilde M_2(\hat Va)\big).
\end{equation}
If $a$ is the density $n$$\times$$n$-matrix, the matrix $\tilde a_2$
is also the density $n$$\times$$n$-matrix, and it has the matrix
elements $(\tilde a_2)_{jn}=(\tilde a_2)_{nj}=0$ for all
$j=1,2,\ldots,n$.

For an example of the qutrit density matrix $a$, the matrix $\tilde
a_2$ reads
\begin{equation}\label{1.14}
\tilde a_2=\left(
          \begin{array}{ccc}
            a_{11}+a_{33} & a_{12} & 0 \\
            a_{21} &  a_{22} & 0\\
            0 & 0 & 0
          \end{array}
        \right).
\end{equation}
In fact, the obtained matrix corresponds to the density
2$\times$2-matrix of the qubit state. The positive map $a\to\tilde
a_2$ provides the map of the qutrit density matrix onto the qubit
density matrix $a\to\widetilde{\tilde a}$ of the form
\begin{equation}\label{1.15}
\widetilde{\tilde a}_2=\left(
          \begin{array}{cc}
            a_{11}+a_{33} & a_{12} \\
            a_{21} &  a_{22}
          \end{array}
        \right).
\end{equation}
The matrix $M_1$ describes another positive map of the qutrit
density matrix
\begin{equation}\label{1.16}
a\to a_1=\hat V^{-1}\big(M_1(\hat V a)\big)=\left(
          \begin{array}{ccc}
            a_{11} & 0 & a_{13} \\
            0 &  a_{22} & 0\\
            a_{31} & 0 &  a_{33}
          \end{array}
        \right).
\end{equation}
This map can also be modified, if one uses the 9$\times$9-matrix
given in the block form
\begin{equation}\label{1.17}
\widetilde M_1=\left(
          \begin{array}{ccc}
            A & B & 0 \\
            0 &  0 & A\\
            0 & 0 &  0          \end{array}
        \right); \qquad A=\left(
          \begin{array}{ccc}
            1 & 0 & 0 \\
            0 &  0 & 1\\
            0 & 0 &  0
          \end{array}
        \right); \qquad B=\left(
          \begin{array}{ccc}
            0 & 1 & 0 \\
            0 &  0 & 0\\
            0 & 0 &  0
          \end{array}
        \right).
\end{equation}
Then one has the positive map
\begin{equation}\label{1.18}
a\to \tilde a_1=\hat V^{-1}\big(\widetilde M_1(\hat V a)\big)=\left(
          \begin{array}{ccc}
            a_{11}+a_{22} & a_{13} & 0 \\
            a_{31} & a_{33} & 0\\
            0 & 0 &  0
          \end{array}
        \right).
\end{equation}
This map can be considered as a map of the qutrit density matrix
onto the qubit density matrix
\begin{equation}\label{1.20}
a\to\widetilde{\tilde a}_1=\left(
          \begin{array}{cc}
            a_{11}+a_{22} & a_{13} \\
            a_{31} &  a_{33}
          \end{array}
        \right).
\end{equation}

Following \cite{PS-MA-VI-1,FP,PSv153,Bregenz,NuovoCim,JRLR2014} it
is easy to check that one has the entropic inequality for the von
Neumann entropies corresponding to the subadditivity condition
\begin{eqnarray}\label{1.21}
-\mbox{Tr}\big\{\big[\hat V^{-1}\big(M_1(\hat
Va)\big)\big]\ln\big[\hat V^{-1}\big(M_1(\hat
Va)\big)\big]\big\}\nonumber\\
-\mbox{Tr}\big\{\big[\hat
V^{-1}\big(M_2(\hat Va)\big)\big]\ln\big[\hat V^{-1}\big(M_2(\hat
Va)\big)\big]\big\}\geq -\mbox{Tr}\{a\ln a\}.
\end{eqnarray}
Using the properties of Tsallis entropy~\cite{Petz2} and the
definition of deformed logarithm
$$\ln_q(x)=\left\{
\begin{array}{c}
\frac{x^{q-1}-1}{q-1},\quad\mbox{if}\quad q\neq 1,\\
\ln x,\quad\mbox{if}\quad q=1,
\end{array}\right.
$$
we obtain the new inequality by replacing $\ln a\to\ln_qa$ in
(\ref{1.21}).

The difference $I_q$ of the left- and right-hand sides of
inequality~(\ref{1.21}) is an analog of the mutual information,
which reflects quantum correlations in the qutrit state. Thus, one
has the information inequality for the qutrit state
\begin{eqnarray}\label{1.21a}
I_q=-\mbox{Tr}\big\{\big[\hat V^{-1}\big(M_1(\hat
Va)\big)\big]\ln\big[\hat V^{-1}\big(M_1(\hat
Va)\big)\big]\big\}\nonumber\\
-\mbox{Tr}\big\{\big[\hat
V^{-1}\big(M_2(\hat Va)\big)\big]\ln\big[\hat V^{-1}\big(M_2(\hat
Va)\big)\big]\big\}+\mbox{Tr}\{a\ln a\}\geq 0.
\end{eqnarray}
We obtain the deformed mutual information by replacing $\ln
a\to\ln_qa$ in (\ref{1.21a}).

The form of inequality~(\ref{1.21a}) is preserved for any
$N$$\times$$N$-matrix $a$. The inequality takes place for an
arbitrary matrix $UaU^\dagger$, where $U$ is a unitary matrix. In
the case where the unitary matrix $U$ is such a matrix that
$UaU^\dagger=a_d$ is a diagonal matrix identified with the
probability distribution $d_1,d_2,d_3$, the inequality yields the
relation valid for the probability distribution of the form
\begin{eqnarray}\label{1.22}
-(d_1+d_2)\ln(d_1+d_2)-d_3\ln d_3-(d_1+d_3)\ln(d_1+d_3)-d_2\ln
d_2\nonumber\\\geq -d_1\ln d_1-d_2\ln d_2-d_3\ln d_3,
\end{eqnarray}
or
\begin{equation}\label{1.23}
-(d_1+d_2)\ln(d_1+d_2)-(d_1+d_3)\ln(d_1+d_3)\ln(d_1+d_3)\geq -d_1\ln
d_1.
\end{equation}

It is easy to see that if we consider the $n$$\times$$n$-matrix $a$
for an arbitrary $n\geq 4$, the map discussed provides the
$(n-1)$$\times$$(n-1)$-matrix $\widetilde{\tilde a}_2$ and
2$\times$2-matrix $\widetilde{\tilde a}_1$
\begin{eqnarray}
\widetilde{\tilde a}_2=\left(
          \begin{array}{cccc}
            a_{1\,1}+a_{n\,n} & a_{1\,2} & \ldots & a_{1\,n-1} \\
 a_{2\,1} & a_{2\,2} & \ldots & a_{2\,n+1} \\
\ldots&\ldots&\ldots&\ldots\\
 a_{n-1\,1}+a_{n-1,\,2} & a_{12} & \ldots & a_{n-1\,n-1}
          \end{array}\right),\label{1.24} \\
          \widetilde{\tilde a}_1=\left(
          \begin{array}{cc}
            a_{1\,1}+a_{2\,2}+\cdots+a_{n-1\,n-1}~~ & ~~a_{1\,n} \\
 a_{n\,1} & ~a_{n\,n}
          \end{array}\right). \label{1.25}
\end{eqnarray}
If the matrix $a$ is the density matrix of a qudit state with
$j=(n-1)/2$, the matrix $\widetilde{\tilde a}_2$ is an analog of the
density matrix of the qudit state with $j=(n-2)/2$, and the matrix
$\widetilde{\tilde a}_1$ is an analog of the density matrix of the
qubit state.

In this case, the subadditivity condition extending
inequality~(\ref{1.21}) reads
\begin{equation}\label{1.26}
-\mbox{Tr}\big(\widetilde{\tilde a}_1\ln \widetilde{\tilde a}_1
\big)-\mbox{Tr}\big(\widetilde{\tilde a}_2\ln \widetilde{\tilde
a}_2\big)\geq-\mbox{Tr}\big(a\ln a\big).
\end{equation}
Deformed inequality is obtained by replacing $\ln a$ with $\ln_qa$.
We should point out that inequality~(\ref{1.26}) takes place for a
single qudit, and the Hilbert space $H$ of the qudit state is not
considered as the tensor-product of the Hilbert spaces of two
physical subsystems. The corresponding matrices of the map $\tilde
M_1$ and  $\tilde M_2$ can be also given in an explicit form.

It is worth pointing out that the other different entropic
inequalities can be obtained for arbitrary density matrices of
composite and non-composite systems if one uses projectors of
different ranks in (\ref{1.3}).

\section{No-signaling property of the single-qudit-state tomogram}

For a composite system, e.g., for a two-qudit state with the density
$N$$\times$$N$-matrix $\rho(1,2)$, the joint tomographic probability
distribution $w(m_1,m_2,u)$ is defined as
\begin{equation}\label{2.1}
w(m_1,m_2,u)=\langle m_1m_2\mid u\rho(1,2)u^\dagger\mid
m_1m_2\rangle.
\end{equation}
The probability distribution, where $m_1=-j_1,-j_1+1,\ldots,j_1$,
$m_2=-j_2,-j_2+1, \ldots, j_2$, and $u$ is a unitary
$N$$\times$$N$-matrix with $N=(2j_1+1)(2j_2+1)$ is the joint
probability distribution of two random spin projections $m_1$ and
$m_2$. The probability distribution depends on the unitary matrix
$u$.

This probability distribution has the property of no-signaling; this
means that for $u=u_1\otimes u_2$, where $u_1$ and $u_2$ are unitary
local transforms in Hilbert spaces of the first and second qudits,
respectively, the marginal probability distributions
\begin{eqnarray}
w_1(m_1,u)=\sum_{m_2=-j_2}^{j_2}w(m_1,m_2,u),\label{2.2}\\
w_2(m_2,u)=\sum_{m_1=-j_1}^{j_1}w(m_1,m_2,u)\label{2.3}
\end{eqnarray}
depend on their own local unitary transforms only; this means that
\begin{eqnarray}
w_1(m_1,u=u_1\times u_2)\equiv w_1(m_1,u_1),\label{2.4}\\
w_2(m_1,u=u_1\times u_2)\equiv w_2(m_2,u_2).\label{2.5}
\end{eqnarray}

Since an arbitrary Hermitian nonnegative  $N$$\times$$N$-matrix
$A=A^\dagger$, such that Tr$\,A=1$, can be considered as the density
matrix of a single qudit, one can formulate an analogous
no-signaling property of such a matrix though it does not describe
the state of a composite system. Thus, one has the property
\begin{eqnarray}
\frac{\partial}{\partial u_2}\sum_{m_2=-j_2}^{j_2}\langle m_1m_2\mid
uAu^\dagger\mid m_1m_2\rangle_{u=u_1\times u_2}=0,\label{2.6}\\
\frac{\partial}{\partial u_1}\sum_{m_1=-j_1}^{j_1}\langle m_1m_2\mid
uAu^\dagger\mid m_1m_2\rangle_{u=u_1\times u_2}=0,\label{2.7}
\end{eqnarray}

If $N\neq(2j_1+1)(2j_2+1)$, one can replace the matrix $A$ in
equalities (\ref{2.6}) and (\ref{2.7}) with the $\tilde
N$$\times$$\tilde N$-matrix $\widetilde A=\left(\begin{array}{cc}
A&0\\
0&0\end{array}\right)$, where $\tilde N=(2j_1+1)(2j_2+1)$. The
matrix $\widetilde A$ is Hermitian, nonnegative, and Tr$\,\widetilde
A=\mbox{Tr}\,A=1$. Then the property of no-signaling given by
equalities (\ref{2.6}) and (\ref{2.7}) is valid for the matrix
$\widetilde A$.

As an example, we consider the 5$\times$5-matrix $\rho$, which is
identified with the density matrix of a qudit with $j=2$, i.e.,
$\rho_{mm'}$, where $m$ and $m'$ are spin projections $-2,-1,0,1,2$.
We construct the 6$\times$6-matrix
$\widetilde\rho=\left(\begin{array}{cc}
\rho & 0\\
0 & 0\end{array}\right)$ and consider this matrix as the density
matrix of a qubit--qutrit system. This means that we use the map for
the basis in the six-dimensional Hilbert space as
\begin{eqnarray*}
\mid -2\rangle\equiv\mid 1/2,1\rangle,~\mid
-1\rangle\equiv\mid 1/2,0\rangle,~
\mid 0\rangle\equiv\mid 1/2,1\rangle,\\
\mid 1\rangle\equiv\mid -1/2,1\rangle,~
\mid 2\rangle\equiv\mid -1/2,0\rangle,~\mid an\rangle\equiv\mid
-1/2,-1\rangle,\end{eqnarray*} where the vector $\mid an\rangle$ is
an additional basis vector in the six-dimensional Hilbert space,
corresponding to zero matrix elements in the matrix
$\widetilde\rho$. Then we can write the no-signaling property in the
form
\begin{eqnarray}
\frac{\partial}{\partial u(2)}\sum_{m_2=-1}^{1}\langle m_1m_2\mid
u(2)\times u(3)\widetilde\rho u^\dagger(2)\times u^\dagger(3)\mid m_1m_2\rangle=0,\label{2.8}\\
\frac{\partial}{\partial u(3)}\sum_{m_1=-1/2}^{1/2}\langle m_1m_2\mid
u(2)\times u(3)\widetilde\rho u^\dagger(2)\times u^\dagger(3)\mid
m_1m_2\rangle=0,\label{2.9}
\end{eqnarray}
where $u(2)$ is the unitary 2$\times$2-matrix and $u(3)$ is the
unitary 3$\times$3-matrix.

In fact, the equalities obtained are new properties of arbitrary
Hermitian nonnegative matrices with unit trace. Analogous
no-signaling properties can be obtained with respect to an arbitrary
product decomposition of unitary $N$$\times$$N$-matrix $u=u_1\times
u_2\times\cdots\times u_m$, if $N=n_1n_2\cdots n_m$.

\section{Monotonicity property of density matrices of a single qudit and two qubits }

It is known that the density matrices of bipartite quantum systems
$\rho(1,2)$ and $\sigma(1,2)$ have the monotonicity
property~\cite{Chuang}, i.e., their relative von Neumann entropies
satisfy the inequality
\begin{equation}\label{3.1}
\mbox{Tr}\,\rho(1,2)\big(\ln\rho(1,2)-\ln\sigma(1,2)\Big)\geq
\mbox{Tr}\,\rho(2)\big(\ln\rho(2)-\ln\sigma(2)\big),
\end{equation}
where $\rho(2)=\mbox{Tr}_1\rho(1,2)$ and
$\sigma(2)=\mbox{Tr}_1\sigma(1,2)$. The density matrices are the
matrices of the density operators $\hat\rho(1,2)$ and
$\hat\sigma(1,2)$ acting in the Hilbert space $H=H_1\times H_2$. The
spaces $H_1$ and $H_2$ are Hilbert spaces of the subsystem states of
the bipartite quantum system. In fact, the monotonicity
property~(\ref{3.1}) can be obtained for an arbitrary
$N$$\times$$N$-matrix $R$ with the properties $R^\dagger=R$,
Tr$\,R=1$, and $R\geq 0$.

The matrix $R$ can be considered, e.g., as the density matrix of a
single qudit state, such that $N=2j+1$; this means that the density
matrix $R$ describes the state of the system, which has no
subsystems.

To obtain the inequality called the monotonicity of the single qudit
state, we apply the portrait method of \cite{Vova,Lupo}. We map the
basic vectors of the Hilbert space $H$ $\mid -j\rangle~\mid
-j+1\rangle\cdots\mid j=1\rangle~\mid j\rangle$ onto vectors $\mid
m_1m_2\rangle$, where $m_1=j_1,-j_1+1,\ldots,j_1-1,j_1$ and
$m_2=j_2,-j_2+1,\ldots,j_2-1,j_2$. Then, in view of the identity of
the density matrix properties with respect to the numerical
relations containing the matrix elements of the matrices, which do
not depend on the tensor-product structure of the Hilbert spaces of
states for bipartite qudit system or single qudit system, we obtain
the inequality under consideration. It was proved~\cite{20prd} that
equality in (\ref{3.1}) takes place if and only if the equality
\begin{equation}\label{3.2}
\ln\rho(1,2)-\ln\sigma(1,2)=\ln\rho(2)-\ln\sigma(2)
\end{equation}
holds.

Now we derive an analog of this condition for the monotonicity
property for a single qudit state. We formulate the main result on
the example of the qudit $(j=3/2)$ state. The density matrices
$\rho^{(3/2)}$ and $\sigma^{(3/2)}$ of the qudit states have the
form
\begin{eqnarray}
\rho^{(3/2)}=
\left(\begin{array}{cccc}
\rho_{3/2~3/2}&\rho_{3/2~1/2}&\rho_{3/2~-1/2}&\rho_{3/2~-3/2} \\
\rho_{1/2~3/2}&\rho_{1/2~1/2}&\rho_{1/2~-1/2}&\rho_{1/2~-3/2} \\
\rho_{-1/2~3/2}&\rho_{-1/2~1/2}&\rho_{-1/2~-1/2}&\rho_{-1/2~-3/2} \\
\rho_{-3/2~3/2}&\rho_{-3/2~3/2}&\rho_{-3/2~-1/2}&\rho_{-3/2~-3/2}
\end{array}\right),\label{3.3}\\
\sigma^{(3/2)}= \left(\begin{array}{cccc}
\sigma_{3/2~3/2}&\sigma_{3/2~1/2}&\sigma_{3/2~-1/2}&\sigma_{3/2~-3/2} \\
\sigma_{1/2~3/2}&\sigma_{1/2~1/2}&\sigma_{1/2~-1/2}&\sigma_{1/2~-3/2} \\
\sigma_{-1/2~3/2}&\sigma_{-1/2~1/2}&\sigma_{-1/2~-1/2}&\sigma_{-1/2~-3/2} \\
\sigma_{-3/2~3/2}&\sigma_{-3/2~3/2}&\sigma_{-3/2~-1/2}&\sigma_{-3/2~-3/2}
\end{array}\right),\label{3.4}
\end{eqnarray}
Then we can write the monotonicity property explicitly
\begin{eqnarray}
&&\mbox{Tr}\left[\rho^{(3/2)}\left(\ln\rho^{(3/2)}-\ln\sigma^{(3/2)}\right)\right]\nonumber\\
&&\geq\mbox{Tr}\left\{\left(\begin{array}{cc}
\rho_{3/2~3/2}+\rho_{-1/2~-1/2}&\rho_{3/2~1/2}+\rho_{1/2~-3/2} \\
\rho_{1/2~3/2}+\rho_{-3/2~-1/2}&\rho_{1/2~1/2}+\rho_{-3/2~-3/2}
\end{array}\right)\right.\nonumber\\
&&\times\left.\left[\ln\left(\begin{array}{cc}
\rho_{3/2~3/2}+\rho_{-1/2~-1/2}&\rho_{3/2~1/2}+\rho_{1/2~-3/2} \\
\rho_{1/2~3/2}+\rho_{-3/2~-1/2}&\rho_{1/2~1/2}+\rho_{-3/2~-3/2}
\end{array}\right)\right.\right.\nonumber\\
&&-\left.\left.\ln\left(\begin{array}{cc}
\sigma_{3/2~3/2}+\sigma_{-1/2~-1/2}&\sigma_{3/2~1/2}+\sigma_{1/2~-3/2} \\
\sigma_{1/2~3/2}+\sigma_{-3/2~-1/2}&\sigma_{1/2~1/2}+\sigma_{-3/2~-3/2}
\end{array}\right)\right]\right\}.\label{3.5}
\end{eqnarray}
Also the inequality holds if one makes all possible permutations of
the indices $3/2,1/2,-1/2,-3/2$ of the matrix elements of matrices
$\rho^{(3/2)}$ and $\sigma^{(3/2)}$.

As the other example, we consider bipartite system of two qubits
with the density matrix $\rho(1,2)$
\begin{eqnarray}\label{3.6}
&&\rho(1,2)=\nonumber\\
&&\left(\begin{array}{cccc}
\rho_{1/2~1/2~1/2~1/2}&\rho_{1/2~1/2~1/2~-1/2}&\rho_{1/2~1/2~-1/2~1/2}&\rho_{1/2~1/2~-1/2~-1/2}\\
\rho_{1/2~-1/2~1/2~1/2}&\rho_{1/2~-1/2~1/2~-1/2}&\rho_{1/2~-1/2~-1/2~1/2}&\rho_{1/2~-1/2~-1/2~-1/2}\\
\rho_{-1/2~1/2~1/2~1/2}&\rho_{-1/2~1/2~1/2~-1/2}&\rho_{-1/2~1/2~-1/2~1/2}&\rho_{-1/2~1/2~-1/2~-1/2}\\
\rho_{-1/2~-1/2~1/2~1/2}&\rho_{-1/2~-1/2~1/2~-1/2}&\rho_{-1/2~-1/2~-1/2~1/2}&\rho_{-1/2~-1/2~-1/2~-1/2}
\end{array}\right)\nonumber\\
\end{eqnarray}
and the density matrix $\sigma(1,2)$ of the same form
\begin{eqnarray}\label{3.7}
&&\sigma(1,2)=\nonumber\\
&& \left(\begin{array}{cccc}
\sigma_{1/2~1/2~1/2~1/2}&\sigma_{1/2~1/2~1/2~-1/2}&\sigma_{1/2~1/2~-1/2~1/2}&\sigma_{1/2~1/2~-1/2~-1/2}\\
\sigma_{1/2~-1/2~1/2~1/2}&\sigma_{1/2~-1/2~1/2~-1/2}&\sigma_{1/2~-1/2~-1/2~1/2}&\sigma_{1/2~-1/2~-1/2~-1/2}\\
\sigma_{-1/2~1/2~1/2~1/2}&\sigma_{-1/2~1/2~1/2~-1/2}&\sigma_{-1/2~1/2~-1/2~1/2}&\sigma_{=1/2~1/2~-1/2~-1/2}\\
\sigma_{-1/2~-1/2~1/2~1/2}&\sigma_{-1/2~-1/2~1/2~-1/2}&\sigma_{-1/2~-1/2~-1/2~1/2}&\sigma_{-1/2~-1/2~-1/2~-1/2}
\end{array}\right).\nonumber\\
\end{eqnarray}
The standard monotonicity property is given by the inequality
\begin{equation}\label{3.8}
\mbox{Tr}\big\{\rho(1,2)\big[\ln\rho(1,2)-\ln\sigma(1,2)\big]\big\}\geq
\mbox{Tr}\big\{\rho(2)\big[\ln\rho(2)-\ln\sigma(2)\big]\big\},
\end{equation}
where
\begin{eqnarray}
&&\mbox{Tr}_1\rho(1,2)=\rho(2)=\nonumber\\
&&\left(\begin{array}{cc}
\rho_{1/2~1/2~1/2~1/2}+\rho_{-1/2~1/2~-1/2~1/2}&\rho_{1/2~1/2~1/2~-1/2}+\rho_{-1/2~1/2~-1/2~-1/2}\\
\rho_{1/2~-1/2~1/2~1/2}+\rho_{-1/2~-1/2~-1/2~1/2}&\rho_{1/2~-1/2~1/2~-1/2}+\rho_{-1/2~-1/2~-1/2~-1/2}
\end{array}\right)\nonumber\\
&&\label{3.9}\\
&&\mbox{Tr}_1\sigma(1,2)=\sigma(2)=
\nonumber\\
&&\left(\begin{array}{cc}
\sigma_{1/2~1/2~1/2~1/2}+\rho_{-1/2~1/2~-1/2~1/2}&\sigma_{1/2~1/2~1/2~-1/2}+\rho_{-1/2~1/2~-1/2~-1/2}\\
\sigma_{1/2~-1/2~1/2~1/2}+\rho_{-1/2~-1/2~-1/2~1/2}&\sigma_{1/2~-1/2~1/2~-1/2}+\rho_{-1/2~-1/2~-1/2~-1/2}
\end{array}\right).\nonumber\\
&&\label{3.9}
\end{eqnarray}
But there exists the other monotonicity inequality of the form
\begin{equation}\label{3.10}
\mbox{Tr}\big\{\rho(1,2)\big[\ln\rho(1,2)-\ln\sigma(1,2)\big]\big\}\geq
\mbox{Tr}\big\{\rho'(2)\big[\ln\rho'(2)-\ln\sigma'(2)\big]\big\},
\end{equation}
where the 2$\times$2-matrix $\rho'(2)$ is not obtained by the
partial trace over degrees of freedom of the first subsystem of the
matrix $\rho(1,2)$. For example, consider the matrices $\rho'(2)$
and $\sigma'(2)$
\begin{eqnarray}
&&\rho'(2)= \nonumber\\
&&\left(\begin{array}{cc}
\rho_{1/2~-1/2~1/2~-1/2}+\rho_{1/2~1/2~1/2~1/2}&\rho_{1/2~-1/2~-1/2~1/2}+\rho_{1/2~1/2~-1/2~-1/2}\\
\rho_{-1/2~1/2~1/2~-1/2}+\rho_{-1/2~-1/2~1/2~1/2}&\rho_{-1/2~1/2~=1/2~1/2}+\rho_{-1/2~-1/2~-1/2~-1/2}
\end{array}\right),\nonumber\\
\label{3.11}\\
&&\sigma'(2)= \nonumber\\
&&\left(\begin{array}{cc}
\sigma_{1/2~-1/2~1/2~-1/2}+\sigma_{1/2~1/2~1/2~1/2}&\sigma_{1/2~-1/2~-1/2~1/2}+\sigma_{1/2~1/2~-1/2~-1/2}\\
\sigma_{-1/2~1/2~1/2~-1/2}+\sigma_{-1/2~-1/2~1/2~1/2}&\sigma_{-1/2~1/2~=1/2~1/2}+\sigma_{-1/2~-1/2~-1/2~-1/2}
\end{array}\right).\nonumber\\
\label{3.12}
\end{eqnarray}

The monotonicity property~(\ref{3.10}) with matrices
$\rho'(2)$~(\ref{3.11}) and $\sigma'(2)$~(\ref{3.12}) reflects
quantum correlations in the system of two qubits. The involved
2$\times$2-matrices $\rho'(2)$ and $\sigma'(2)$ are not the density
matrices of the first and second qubits. These matrices contain
contributions of the density matrices of the both subsystems.

One can present other matrices analogous but different from matrices
$\rho'(2)$ and $\sigma'(2)$ and different from the standard density
matrices $\rho(2)$ and $\sigma(2)$, which satisfy the monotonicity
inequality. If due to the property~(\ref{3.8}), the inequality
converts to the equality, all the matrices $\rho(2)$, $\sigma(2)$,
$\rho'(2)$, $\sigma'(2)$, etc. satisfy equality~(\ref{3.2})
\begin{equation}\label{3.13}
\big[\ln\rho(2)-\ln\sigma(2)\big]=
\big[\ln\rho'(2)-\ln\sigma'(2)\big].
\end{equation}

\section{Bell inequalities and their violation for states of a qudit
with $\mathbf{j=3/2}$}

The developed approach provides the possibility to extend the
discussion of Bell inequalities, e.g., considered in \cite{CHSH} for
two qubits to the case of a qudit with $j=3/2$ (as it was mentioned
in \cite{Vova}).

Since purely numerical properties of matrices (which are tables
containing real or complex numbers) do not depend on the structure
of Hilbert spaces and operators acting in these Hilbert spaces
having the matrix representations under consideration, one can
obtain relations for the functions of the matrix elements, which are
universal and valid for many systems represented by the numerical
matrices.

We demonstrate the known example of such a numerical
4$\times$4-matrix
\begin{equation}\label{B}
\rho=\dfrac{1}{2}\left(\begin{array}{cccc}
1 & 0 & 0 & 1 \\
0 & 0 & 0 & 0 \\
0 & 0 & 0 & 0 \\
1 & 0 & 0 & 1
\end{array}\right).
\end{equation}
If this matrix is identified with the density matrix of the state
with density operator  $\hat\rho(1,2)$ of two qubits, where the
basis in the Hilbert space $H=H_1\times H_2$ has the product form
$\mid m_1m_2\rangle=\mid m_1\rangle\mid m_2\rangle$ with spin
projection $m_{1,2}=\pm 1/2$, the matrix $\rho$~(\ref{B})
corresponds to the pure state
\begin{equation}\label{B2}
\mid\psi\rangle=\frac{1}{\sqrt 2}\,\big(\mid 1/2~1/2\rangle+\mid -1/2~-1/2\rangle
\big).
\end{equation}

On the other hand, one can consider the matrix $\rho$ as the density
matrix of the state $\mid\varphi\rangle$ of a single qudit with
$j=3/2$. In this case, the state $\mid\varphi\rangle$ is the vector
in the Hilbert space $H_{3/2}$, and there is no decomposition of
this space into ``natural'' product of two Hilbert spaces.

We use the basis vectors $\mid m\rangle$, where
$m=-3/2,-1/2,1/2,3/2$, to construct the matrix $\rho$~(\ref{B})
containing the matrix elements of the density operator $\hat\rho$ as
$\rho_{mm'}=\langle m\mid\hat\rho\mid m'\rangle$. The matrix
$\rho$~(\ref{B}) corresponds to the pure state $\mid\varphi\rangle$
of a single qudit with $j=3/2$, which is a linear superposition of
two states
\begin{equation}\label{B3}
\mid\varphi\rangle=\frac{1}{\sqrt 2}\,\big(\mid 3/2\rangle+\mid -3/2\rangle
\big).
\end{equation}

From the viewpoint of reflecting the physical properties of the
systems, the operators $\hat\rho(1,2)$ and $\hat\rho$ and the
Hilbert spaces $H=H_1\times H_2$ and $H_{3/2}$ are quite different,
but the numerical matrix $\rho$ given by (\ref{B}) is universal and
the same for the two different physical situations. In view of this
observation, we discuss the properties which depend only on the
numerical properties of the matrix $\rho$.

The entanglement phenomenon can be associated with the property of
the matrix $\rho$~(\ref{B}) represented in a form of the convex sum
\begin{equation}\label{B4}
\rho=\sum_kp_k\rho^{(k)}_1\otimes\rho^{(k)}_2,\qquad p_k\geq
0,\qquad\sum p_k=1,
\end{equation}
where Hermitian nonnegative matrices $\rho^{(k)}_1$ and
$\rho^{(k)}_2$ are 2$\times$2-matrices and
Tr$\,\rho^{(k)}_1=\mbox{Tr}\,\rho^{(k)}_2=1$. The Peres--Horodecki
{\it ppt}-criterion can be formulated for the matrix $\rho$ without
using the local structure of the states in Hilbert spaces $H$ or
$H_{3/2}$. It is obvious that the necessary condition for the
existence of the decomposition of numerical matrix $\rho$~(\ref{B4})
written in a generic form
\begin{equation}\label{B5}
\rho= \left(\begin{array}{cccc}
\rho_{11} & \rho_{12} & \rho_{13} & \rho_{14}\\
\rho_{21} & \rho_{22} & \rho_{23} & \rho_{24}\\
\rho_{31} & \rho_{32} & \rho_{33} & \rho_{34}\\
\rho_{41} & \rho_{42} & \rho_{43} & \rho_{44}
\end{array}\right),
\end{equation}
where the matrix elements $\rho_{jk}$ $(j,k=1,2,3,4)$ are complex
numbers, is given by the following statement. If the numerical
matrix $\rho$~(\ref{B5}) can be presented in the form~(\ref{B4}),
one has the consequence that the matrix $\widetilde\rho=
\left(\begin{array}{cccc}
\rho_{11} & \rho_{21} & \rho_{13} & \rho_{23}\\
\rho_{12} & \rho_{22} & \rho_{14} & \rho_{24}\\
\rho_{31} & \rho_{41} & \rho_{33} & \rho_{43}\\
\rho_{32} & \rho_{42} & \rho_{34} & \rho_{44}
\end{array}\right)$ is nonnegative Hermitian matrix with
Tr$\,\widehat\rho=1$. If the matrix $\rho$ is considered as the
density matrix of two qubits, i.e., is identified with numerical
matrix $\langle m_1m_2\mid\hat\rho(1,2)\mid m'_1m'_2\rangle$, the
formulated statement is the {\it ppt}-criterion.

The statement formulated is valid also in the case where the matrix
$\rho$ is identified with the density matrix of the qudit state with
$j=3/2$.

For the matrix $\rho$~(\ref{B5}), one can formally introduce a
unitary tomographic probability distribution
\begin{equation}\label{B6}
w_n(u)=\langle n\mid u\rho u^\dagger\mid n\rangle,\qquad n=1,2,3,4,
\end{equation}
where $u$ is the unitary 4$\times$4-matrix.

In the case where $\rho$ is identified with the density matrix
$\rho_{m_1m_2m'_1m'_2}$, the distribution $w_{m_1m_2}(u)$ is the
unitary tomogram of the two-qubit state.

In the case where the matrix $\rho$ is identified with the density
matrix of a single qudit state with $j=3/2$, the
distribution~(\ref{B6}) is identified with the unitary tomogram
$w_n(u)$ of the spin-3/2 state.

Numerically both distributions $w_m(u)$ and $w_{m_1m_2}(u)$
coincide, but the tomograms have different physical interpretations.

For the case $u=u_1\times u_2$, where $u_1$ and $u_2$ are the
spin-1/2 representation matrices depending on the Euler angles $(
\varphi,\theta)$, which provide the unit vector $\vec
n=(\sin\theta\cos\varphi,\sin\theta\sin\varphi,\cos\theta)$, the
tomogram $w_{m_1m_2}(\vec n_1,\vec n_2)$ is the joint probability
distribution of two random spin-1/2 projections $m_1$ and $m_2$ onto
the corresponding quantization directions $\vec n_1$ and $\vec n_2$.

In the case of $\rho$ identified with the $j=3/2$-qudit state, the
tomogram $w_m(u)$, where $u$ is the matrix of unitary irreducible
representation, converts to the tomogram $w_m(\vec n)$, which is the
probability to have one random spin-3/2 projection $m$ onto the
quantization axis $\vec n$.

We point out that both tomograms $w_m(u)$ and $w_{m_1m_2}(u)$ are
the particular values of the distribution $w_n(u)$, where one
chooses the particular subgroups of the unitary $u(4)$ group and
uses various maps of matrix element indices.

For the two-qubit case, one uses the bijective map
$$~1/2~1/2\leftrightarrow 1,~~1/2~-1/2\leftrightarrow 2,~
~-1/2~1/2\leftrightarrow 3,~~-1/2~-1/2\leftrightarrow 4.~$$

For the $j=3/2$-qudit case, one uses the bijective map
$$~3/2\leftrightarrow 1,~~1/2\leftrightarrow 2,~
~-1/2\leftrightarrow 3,~~-3/2\leftrightarrow 4.~$$

Now we formulate an extension of the Bell inequality (in fact, CHSH
inequality) known for the two-qubit case in the tomographic form to
be used to apply the approach to obtain the inequality for an
arbitrary matrix $\rho$~(\ref{B5}). Also we obtain the inequality
for the $j=3/2$-qudit state.

The general approach uses the tomogram $w_n(u)$.

We construct the stochastic 4$\times$4-matrix $M$ for which four
columns are the probability vectors $\vec w(u_k)$, $k=1,2,3,4$ with
four components $w_n(u)$. If the matrix $\rho$ has the
structure~(\ref{B4}), the probability vector $\vec w(u)$ is also of
a special form. To demonstrate the inequality, first we consider
4$\times$4-matrix $M$ of the form of the tensor product of two
stochastic 2$\times$2-matrices
\begin{equation}\label{B7}
M=\left(
          \begin{array}{cc}
            x & y \\
            1-x & 1-y
          \end{array}
        \right)\otimes\left(
          \begin{array}{cc}
            z & t \\
            1-z & 1-t
          \end{array}
        \right),\qquad 0\leq x,y,z,t\leq 1.
\end{equation}
Then, it can be checked that the function
$B(x,y,z,t)=\mbox{Tr}\,(IM)$, where $I=\left(
          \begin{array}{cccc}
            1 & -1 & -1 & 1\\
 1 & -1 & -1 & 1\\
 1 & -1 & -1 & 1\\
 -1 & 1 & 1 & -1          \end{array}
        \right)$ satisfies the Laplace equation
\begin{equation}\label{B8}
\left(\frac{\partial^2}{\partial x^2}+\frac{\partial^2}{\partial y^2}+
\frac{\partial^2}{\partial z^2}+\frac{\partial^2}{\partial
t^2}\right)B(x,y,z.t)=0.
\end{equation}
This means that the function $B(x,y,z,t)$ has the maximum and
minimum values on the boundary of the cube in the four-dimensional
space. Thus, taking values of the function $B(x,y,z,t)$ for
arguments zero or one, we obtain the inequality
\begin{equation}\label{B9}
|B(x,y,z,t)|\leq 2,
\end{equation}
which is easy to check. Using the inequality for the convex sum
$|\sum_kp_kz_k|\leq\sum_k|z_k|$, we obtain the inequality for the
matrix $M=\sum_kp_kM_k$, where $M_k$ is given by (\ref{B7}) with
$x,y,z,t$ replaced by $x_k,y_k,z_k,t_k$ and $0\leq p_k\leq 1$. The
inequality for this matrix reads
\begin{equation}\label{B10}
|\mbox{Tr}\,(IM)|\leq 2.
\end{equation}

Now we consider the 4$\times$4-matrix using four vectors $\vec
w(u_k)$, i.e.,
\begin{equation}\label{B11}
M(u_1u_2u_3u_4)= \left(\begin{array}{cccc}
w_1(u_1) & w_1(u_2) & w_1(u_3) & w_1(u_4)\\
w_2(u_1) & w_2(u_2) & w_2(u_3) & w_2(u_4)\\
w_3(u_1) & w_3(u_2) & w_3(u_3) & w_3(u_4)\\
w_4(u_1) & w_4(u_2) & w_4(u_3) & w_4(u_4)
\end{array}\right).
\end{equation}
The matrix $M$ is the stochastic matrix. We choose specific matrices
$u_k$ of the product form
\begin{equation}\label{B12}
u_1=u_a\otimes u_b,\quad u_2=u_a\otimes u_c,\quad u_3=u_d\otimes
u_a,\quad u_4=u_d\otimes u_c,
\end{equation}
where the matrices $u_a$, $u_b$, $u_c$, and $u_d$ are arbitrary
unitary 2$\times$2-matrices. In this particular case, one can
construct the number
\begin{equation}\label{B13}
B(u_a,u_b,u_c,u_d)=\mbox{Tr}\,\big(IM(u_1,u_2,u_3,u_4)\big).
\end{equation}
If the 4$\times$4-matrix $\rho$ has the structure~(\ref{B4}), the
number $B(u_a,u_b,u_c,u_d)$~(\ref{B13}) satisfies the inequality
\begin{equation}\label{B14}
|B(u_a,u_b,u_c,u_d)|\leq 2.
\end{equation}
It is an analog of the CHSH~\cite{CHSH} inequality proved in view of
generic inequality~(\ref{B9}). Inequality~(\ref{B14}) is valid
independently of any product structure of the Hilbert space.

If the matrix $\rho$ can be presented in the form~(\ref{B4}), one
has an extended Bell inequality~(\ref{B14}) for arbitrary unitary
2$\times$2-matrices $u_a$, $u_b$, $u_c$, and $u_d$. From this
observation follows the statement that for the density matrix of
qudit with $j=3/2$ of the form~(\ref{B4}) one has the Bell
inequality. We call a such density matrix of the qudit state the
separable matrix. If the inequality is violated, it is easy to show
that one has the Cerelson bound $2\sqrt 2$~\cite{Cirelson} for the
number $B(u_a,u_b,u_c,u_d)$ calculated for a single qudit state with
$j=3/2$.

The physical interpretation of the Bell inequality and its violation
introduced for $j=3/2$ is different from the standard case of two
qubits. For two qubits, quantum correlations exist for the degrees
of freedom of two different subsystems of qubits. For a qudit with
$j=3/2$, quantum correlations exist for degrees of freedom (spin
projections) of the same system. The violation of Bell inequalities
for two qubits was checked experimentally~\cite{Aspect}, and one can
try to check the violation of Bell inequalities for qudit with
$j=3/2$.

\section{Conclusions}
To conclude, we list the main results of our work.

We showed that quantum correlations of the qudit systems expressed
in terms of the inequalities for the system density matrices, such
as the subadditivity condition and Bell inequalities (known for
composite systems), exist also for non-composite systems (like a
single qudit). This means that the inequalities for densities
matrices, in fact, are universal inequalities for arbitrary
nonnegative Hermitian matrices with unit trace.

The quantum correlations of composite quantum systems like systems
of $N$ qubits are known to a provide the resource for quantum
technologies, e.g., for quantum computing. Since analogous quantum
correlations exist in the non-composite systems like single qudits,
there is a possibility to use them as a resource for quantum
technologies as well. For example, the density matrix of the $N$
qubit system and the density matrix of qudit with $j=(N-1)/2$
contain an equivalent resource of the quantum correlations. The
problem of finding the possibility to use this resource needs an
extra clarification.

It is worth pointing out that analogs of particular matrix
inequalities, obtained here, can be extended using other projectors
in formula~(\ref{1.3}) for positive maps. Also the deformed matrix
inequalities expressed in terms of $q$-logarithm can be obtained for
density $N$$\times$$N$-matrices of arbitrary composite and
non-composite quantum systems of qudits, and this will be done in
the future publication.

\section*{Acknowledgements}
The authors thank the Organizers of the Workshop ad Memoriam of
Carlo Novero ``Advances in Foundations of Quantum Mechanics and
Quantum Information with Atoms and Photons'' and especially Prof.
Marco Genovese for invitation and kind hospitality.

\end{document}